# A study of volatile compounds in the breath of children with type 1 diabetes

S Stevens[1], C Garner[2], C Wei[3], R Greenwood[4], J Hamilton-Shield[5], B de Lacy Costello[1], N Ratcliffe[1] and C Probert[2,6]

[1]University of the West of England, Institute of Biosensing Technology, Frenchay campus, Coldharbour lane, Bristol, BS16 1QY
[2]University of Bristol, Level 7, Bristol Royal Infirmary, Marlborough Street, Bristol, BS2 8HW
[3]Bristol Royal Hospital for Children, Paul O'Gorman Building, Upper Maudlin Street, Bristol, BS2 8BJ
[4]Department of Research and Development, Bristol Royal Infirmary, Bristol, UK BS2 8HW
[5] NIHR Bristol Biomedical Research Unit in Nutrition
[6]Department of Gastroenterology, Institute of Translational Medicine, University of Liverpool, Crown Street, L69 3GE, UK

E-mail: Ben.deLacyCostello@uwe.ac.uk

**Short title** Breath analysis of children with type 1 diabetes

**Abstract**: A pilot study of exhaled volatile compounds and their correlation with blood glucose levels in eight children with type 1 diabetes is reported. Five paired blood and breath samples were obtained from each child over a 6 hour period. The blood glucose concentration ranged from 41.4 to 435.6 mg/dL. Breath samples were collected in Tedlar® bags and immediately evacuated through thermal desorption tubes packed with Carbopack B and C. The VOCs were later recovered by thermal desorption and analysed using gas chromatography mass spectrometry. The study identified 74 volatile compounds present in at least 10% of the patient samples. Of these 74 volatiles 36 were found in all patient samples tested. Further analysis of the 36 compounds found that none showed significant overall correlation with blood glucose levels. Isoprene showed a weak negative correlation with blood glucose levels. Acetone was found to have no correlation with blood glucose levels for the patients studied. Some patients showed significant individual correlation between the relative peak areas of certain compounds and blood glucose levels. However, there was no consistent pattern observed within these results across all 8 patients.
Additional breath samples were collected in Tedlar® bags and analysed using SIFT-MS for 3 of the patients and a healthy control. The levels of 24 volatiles are reported and were found to be generally consistent with previously reported SIFT-MS data. In agreement with the GCMS data, no compounds exhibited a significant overall correlation with blood glucose level. Acetone exhibited a negative correlation with blood glucose but this could largely be attributed to higher levels in the control versus the patient group. Again, some patients and the control exhibited individual correlations between the concentration of VOCs and blood glucose levels. However, the findings were not completely consistent with the GCMS data.
This study found a lack of correlation between breath VOCs and blood glucose levels using two methods. Further work involving larger numbers of diabetic children stratified according to age, sex and disease duration/ history would be required in order to confirm these findings.

# 1. Introduction

Type 1 diabetes is a major health problem. The current prevalence of Type 1 diabetes in children in the UK is one per 700–1,000 [diabetes.org.uk]. Good glycaemic control, by regular blood glucose monitoring, is the cornerstone of type 1 diabetes management. It reduces the risk of long-term micro- and macro-vascular complications [White *et al.* 2001] and helps to prevent hypoglycaemia. Self-monitoring protocols are vital to improved long-term control in childhood diabetes [Rosilio *et al.* 1998 and Craig *et al.* 2007]. However, this is hampered by poor adherence [Hermansson et al. 1986], in part due to pain from lancet sampling to test blood glucose [Burge *et al.* 2001]. Needle phobia has a prevalence of at least 10% in the general population [Hamilton, 1995], but in children with diabetes the prevalence may be as high as 27% [Simmons *et al.* 2007]. Attempts to reduce pain by making shallower puncture sites had a limited impact on patient acceptability [Pecaud *et al.* 1999]. Needleless devices have been trialled but long-term acceptance is poor as a result of inconvenience, pain and skin irritation [Weinzimer *et al.* 2009]. Much effort has been focussed on the development of alternative methods of continuous glucose monitoring [Ferrante do Amaral and Wolf 2008] but they are not being used to replace self-monitoring in children. A non-invasive, non-traumatic test for estimating blood glucose concentration would prove more acceptable to diabetics, particularly children.

The measurement of exhaled volatile compounds could have many advantages over current blood glucose monitoring methods, if a marker compound that is predictive of blood glucose could be identified. Acetone [Wang et al. 2010], ethanol (in combination with acetone) [Galassetti *et al.* 2005], carbon monoxide [Paredi *et al.* 1999] methyl nitrate [Novak *et al.* 2007], C4–C20 alkanes and monomethylated alkanes [Phillips *et al.* 2004] have each been reported to correlate with blood glucose in adults.

Wang *et al.* [Wang *et al.* 2010] utilised cavity ringdown spectroscopy to analyse the breath of 34 patients with type 1 diabetes, 10 with type 2 diabetes and 15 healthy controls; they found no overall correlation between breath VOCs and blood glucose or blood glycohaemoglobin A1c (HbA1c) in type 1 diabetics. Deng *et al.* developed a solid-phase micro-extraction (SPME) method involving an on fibre derivatisation of acetone to acetone oxime for the accurate determination of breath acetone levels [Deng *et al.* 2004]; they used the method to measure the breath acetone concentrations of 15 patients with type 2 diabetes and 15 controls but did not link the analysis to blood glucose levels.

Selected ion flow tube - Mass spectrometry (SIFT-MS), due to its direct sampling capabilities and accurate quantification selected ion flow tube, has been used extensively for the measurement of breath acetone and other VOCs in both healthy individuals [Dummer *et al.* 2010 and Spanel *et al.* 2007] and patients with type 1 [Turner *et al.* 2009] and type 2 [Storer *et al.* 2011] diabetes. Proton-transfer mass spectrometry (PTRMS) has been used to measure the breath acetone concentrations in 243 healthy volunteers [Schwarz *et al.* 2009].

This report concerns a GC-MS and SIFT-MS study of exhaled volatile compounds from the breath of children with type 1 diabetes. The measured levels of these volatiles were correlated with blood glucose levels to assess their potential as non-invasive markers of blood glucose concentration in this age group.

# 2. Experimental

## 2.1. Patients and approval

Eight children (aged 6 to 16 years) (Table 1) with type 1 diabetes were invited to participate in the study. Ethical approval was granted by Salisbury and Wiltshire Research Ethics Committee and clinical research governance approval and sponsorship granted by the host institution. An age matched healthy control was also recruited (age 11 years). Informed consent was obtained from the parents of participants.

**Table 1.** Age and gender of patients and control.

| Sex/ID | Age | | BMI | Insulin regime | Insulin dose/kg | Duration of diabetes (years) | HbA1c (mmol/mol) |
|---|---|---|---|---|---|---|---|
| | Years | Months | | | | | |
| F 1 | 16 | 4 | 32.4 | Insulin pump | 1.1 | 8.8 | 58 |
| F 2 | 15 | 5 | 24.1 | Basal/bolus | 1.5 | 8.5 | n/a* |
| F 3 | 14 | 6 | 20.1 | Basal/bolus | 1.0 | 1.9 | 52 |
| M 1 | 13 | 10 | 19.7 | Basal/bolus | 1.0 | 5.4 | 60 |
| M 2 | 13 | 0 | 17.9 | Basal/bolus | 0.9 | 8.6 | 48 |
| M 3 | 11 | 10 | 17.0 | Basal/bolus | 1.0 | 0.5 | 72 |
| M 4 | 10 | 10 | 22.2 | Basal/bolus | 1.3 | 4.3 | 68 |
| M 5 | 6 | 6 | 15.0 | BD mixed insulin | 0.6 | 1.8 | 54 |

*this patient is Hb Punjabi (fructosamine level 317 micromol/l) equivalent to HbA1c of 53mmol/mol

*2.2. Clinical procedures*

Clinical procedures were carried out in the clinical research unit at the Bristol Royal Hospital for Children. Serial blood samples were drawn at 90-minute intervals, via an intravenous cannula. Glucose concentration was determined using the hexokinase method [Tietz 1995] in the clinical chemistry laboratory of the hospital. The blood glucose of each participant was permitted to vary according to his or her usual self-management: snack refreshments were provided at 10:00 and 14.30 (sugar-free squash, apples, crisps and biscuits) and lunch was provided at 12:00.

*2.3 Breath VOC collection*

The children had breakfast prior to attending the research unit. A blood and breath sample was collected from each child at 5 points during the day (Table 2). The samples were paired, immediately following the collection of a blood sample a breath sample was also collected. Breath was collected by exhalation through a PTFE mouthpiece into a 3 litre Tedlar® bag (Adtech, UK) until it was full, then the bag was closed. Each child was asked to inhale and then exhale fully into the bag to provide a full vital capacity sample. Turner et al. (2009) found that measured concentrations of acetone in whole breath using a similar method were close to those measured in alveolar breath. It should be noted that for certain compounds it is known that the route of expiration can affect the concentration. For example it was found that ammonia, ethanol and hydrogen cyanide concentrations were significantly lower in nose-exhaled breath compared to mouth-exhaled breath [Wang *et al*. 2008]. We adopted the sampling methodology of a full vital capacity sample via the mouth as it was easily tolerated by the patients and enabled collection of the required volume in a short time period.

The bags were flushed with scrubbed nitrogen 3 times to remove any contamination prior to their use. A control sample was taken using nitrogen gas stored in a Tedlar® bag on each study day to assess contamination; this showed the presence of phenol and N,N-dimethylacetamide on each occasion, as expected [Steeghs *et al*. 2007]. To minimise the potential loss of volatiles by partitioning [Groves and Zellers 1996], all samples were evacuated from the bags, onto collection tubes, immediately following their collection.

The collection tubes (PerkinElmer, UK) were empty glass tubes packed with Carbopack B and C (Sigma Alrich, UK) adsorbent, selected for their ability to reversibly trap a wide range of VOCs. These were used with the PerkinElmer TurboMatrix TD 50 automated thermal desorption (ATD) system. Each tube was conditioned using the ATD system at a temperature of 350°C and run via GCMS prior to use to test for contaminants. During transportation to and from the hospital the clean tubes were sealed to air using brass Swagelock® storage caps with PTFE ferrules.

The breath filled Tedlar® bag was connected via a clean collection tube to an in-house built bag evacuation device designed to evacuate 2 litres of gas through the collection tube at a constant flow rate of 350 ml min$^{-1}$ in order to pre-concentrate the VOCs. The tube was then re-sealed with the Swagelok® storage caps until analysis by GC-MS. All samples were analysed within 24 hours of collection. Room air control samples, collected directly onto collection tubes were also obtained.

An additional breath sample, in a separate Tedlar® bag, was obtained from three patients (F1, M5 and M4) for analysis by SIFT-MS. These bags were taken to the laboratory and analysed on the same day.

*2.4. GC-MS method*

The analyses were undertaken on a Perkin Elmer Clarus 500 quadrupole GC/MS (Perkin Elmer, Beaconsfield, UK) coupled to a TurboMatrix TD50 thermal desorption unit. Pure helium carrier gas of 99.9995% (BOC, Guildford, UK) was passed through a helium purification system, ExcelasorbTM (Supelco, Poole, UK) at 1.1 ml min$^{-1}$. The GC column was a 60 metre long Zebron ZB-624 capillary GC column with an inner diameter of 0.25 mm and a film thickness of 1.4 μm. The system parameters were set as follows:

ATD
| | | |
|---|---|---|
| Carrier gas | – | helium (scrubbed) |
| Head pressure | – | 44.6 psi |
| Reverse dry purge | – | 10 minutes at 90 ml min$^{-1}$ with helium |
| Primary desorption | – | 280°C for 20 minutes at 60 ml min$^{-1}$ with helium |
| Inlet Split | – | off |
| Outlet split | – | off |
| Secondary desorption | – | 310°C for 4 minutes |
| Transfer line | – | 250°C |

GC
| | | |
|---|---|---|
| Initial temperature | – | 35°C with a 1 minute hold |
| Temperature ramp 1 | – | 7°C min$^{-1}$ to 100°C with no hold |
| Temperature ramp 2 | – | 4°C min$^{-1}$ to 200°C with no hold |
| Temperature ramp 3 | – | 20°C min$^{-1}$ to 220°C with 12 minute hold |

MS
| | | |
|---|---|---|
| Initial delay | – | 3.5 minutes |
| Ionisation mode | – | EI, 70eV |
| Scan Mode | – | 10–300 m/z |
| Scan time | – | 0.3 sec |
| Dwell time | – | 0.05 sec |

*2.5 SIFT-MS method*

The analysis was undertaken using a SIFT-MS Voice 200 instrument (Syft Technologies Ltd. Christchurch, NZ). For a detailed description of the Voice 200 instrument and operation refer to Prince *et al.* 2010. For a detailed description of SIFT-MS technology refer to [Smith and Spanel 2005]. Precursor $H_3O^+$, $NO^+$ and $O_2^+$ ions are generated from a microwave air./water discharge at 0.5 Torr. A specific mass is selected by the upstream quadrupole mass filter and then the selected ionic species are injected into fast flowing helium carrier gas in a flow tube reactor. The sample is introduced to the flow tube and measurement of the counts per second of the precursor ions and the resulting product ions is performed via the downstream quadrupole mass filter and ion counting system (m/z range 10-300 amu).The SIFT-MS instrument was used in multiple ion monitoring mode (MIM) for these experiments, the count rate of selected ions of interest over a pre-defined time period being utilised for real time quantification of certain pre-selected compounds. Breath samples collected in Tedlar® bags from 3 patients with type 1 diabetes and a healthy control were introduced into the flow tube of the SIFT-MS at a controlled flow rate (10sccm) via a heated (373K) variable leak valve at the sample inlet. Each sample bag was attached directly to the SIFT-MS sampling line by pushing the connection needle through a septum located on the bag's spigot valve. All samples were analysed within 6 hours of being produced by the patients. The samples were analysed in the order they were produced by the patients. Once connected to the sample line four discrete samples were taken from each sample bag, each of these was designed to detect and quantify specific groups of compounds. The masses of the ion fragments used to quantify each compound are listed after the compound name. Where two fragments of the same mass were used to quantify a given compound then the relevant precursor ion is also noted. The groups were made up of the following compounds:

Group A – heptane (m/z 60 ($C_2H_5NOH^+$), 74 ($C_3H_7NOH^+$), 99 ($C_7H_{15}^+$) (heptane/$NO^+$), 99 ($C_7H_{15}^+$) (heptane/H3O$^+$), 119 ($H_3O^+.C_7H_{16}$)), acetone (m/z 59 ($C_3H_7O^+$)), 1-butanol (m/z 73 ($C_4H_9O^+$)) , benzene (m/z 78 ($C_6H_6^+$) (benzene;$O_2^+$), 78 ($C_6H_6^+$) (benzene;$NO^+$), 108 ($NO.C_6H_6^+$)), ammonia (m/z 17 ($NH_3^+$), 18 ($NH_4^+$)), isoprene (m/z 53 ($C_4H_5^+$), 67 ($C_5H_7^+$), 68 ($C_5H_8^+$) (isoprene;$O_2^+$), 68 ($C_5H_8^+$) (isoprene;$NO^+$), 69 ($C_5H_8.H^+$)), hexane (m/z 105 ($H_3O^+.C_6H_{14}$)), acetaldehyde (m/z 32 ($C_2H_4O^+$), 43 ($CH_3CO^+$), 45 ($C_2H_5O^+$))

Group B – butanone (m/z 72 ($C_4H_8O^+$), 102 ($NO^+.C_4H_8O$)), 1-propanol (m/z 42 ($C_3H_6^+$), 31 ($CH_3O^+$), 59 ($C_3H_7O^+$), 43 ($C_3H_7^+$), 61 ($C_3H_9O^+$)), toluene (m/z 92 ($C_7H_8^+$) (toluene;$O_2^+$), 92 ($C_7H_8^+$) (toluene;$NO^+$), 93 ($C_7H_8.H^+$)), propane (m/z 28 ($C_2H_4^+$), 29 ($C_2H_5^+$), 44 ($C_3H_8^+$), 43 ($C_3H_7^+$)), butanoic acid (m/z 73 ($C_2H_4COOH^+$), 88 ($C_3H_7COOH^+$), 60 ($CH_3COOH^+$), 71 ($C_3H_7CO^+$) (butanoic acid;$NO^+$), 118 ($NO^+.C_3H_7COOH$), 71 ($C_3H_7CO^+$) (butanoic acid;$H_3O^+$), 89 ($C_3H_7COOH_2^+$)), nitric oxide (m/z 30 $NO^+$)), 2-heptanone (m/z 58 ($C_3H_6O^+$), 71 ($C_5H_{11}^+$), 114 $C_7H_{14}O^+$, 144 ($C_7H_{14}O.NO^+$), 115 ($C_7H_{14}OH^+$))

Group C – β-pinene (m/z 107 ($C_8H_{11}^+$)), heptanal (m/z 113 ($C_7H_{13}O^+$), 115($C_7H_{15}O^+$)), hexanal (m/z 72 ($C_4H_8O^+$), 99 ($C_6H_{11}O^+$), (101 ($C_6H_{13}O^+$), methanol (m/z 33 ($CH_5O^+$)), nonane (m/z 147 ($C_9H_{20}.H_3O^+$)), 2-pentanone (m/z 116 ($NO^+.C_5H_{10}O$)), acetic acid (m/z 60 ($CH_3COOH^+$), 90 ($NO^+.CH_3COOH$), 61($CH_3COOH_2^+$) and p-xylene (m/z 91($C_7H_7^+$), 106($C_8H_{10}^+$))

Group D – ethanol (m/z 45 ($C_2H_5O^+$) (ethanol;$O_2^+$), 46 ($C_2H_6O^+$ (ethanol;$O_2^+$), 45 ($C_2H_5O^+$) (ethanol;$NO^+$), 47 ($C_2H_7O^+$)).

The compounds in each group were quantified over a 60 second period. The grouping of the compounds in this way was undertaken to minimise overlap between the selected ions of target compounds. This negates certain potential ionisation conflicts which could adversely affect the accurate quantification of the target compounds. The list of the 24 pre-selected compounds and their concentration (ppb) in the patient samples is displayed in Table 5.

### 3. Results and Discussion

*3.1 Blood glucose*

The blood glucose concentration demonstrated a wide degree of both subject and group variation with a total range of between 41.4 and 435.6 mg/dL during the study (Table 2). The HbA1c results displayed in Table 1 show that all subjects fall within the good to moderate glycaemic control groups. (Range of values 48-72 mmol/mol, which equates to blood glucose levels of 140.4-203.4mg/dL). The duration of disease spanned 0.5-8.8 years. The patients had an average BMI of 21 (s.d 5.4), the majority of patients were in the healthy weight range.

**Table 2.** Time of blood glucose measurement and the result for individual patients

| Time of blood glucose (mins) | Control mg/dL | F1 mg/dL | F2 mg/dL | F3 mg/dL | M1 mg/dL | M2 mg/dL | M3 mg/dL | M4 mg/dL | M5 mg/dL |
|---|---|---|---|---|---|---|---|---|---|
| 0 | 70.2 | 180.0 | 185.4 | 165.6 | 126.0 | 183.6 | 131.4 | 340.2 | 297 |
| 90 | 68.4 | 288.0 | 57.6 | 102.6 | 198 | 64.8 | 75.6 | 226.8 | 226.8 |
| 180 | 91.8 | 309.6 | 41.4 | 91.8 | 158.4 | 90.0 | 59.4 | 187.2 | 122.4 |
| 270 | 86.4 | 189 | 147.6 | 88.2 | 190.8 | 136.8 | 48.6 | 361.8 | 360 |
| 360 | 88.2 | 154.8 | 52.2 | 115.2 | 126 | 145.8 | 68.4 | 300.6 | 435.6 |

*3.2 GC-MS*

Identification of the chromatographic peaks was carried out through spectral matching using the National Institute of Standards and Technology (NIST) Library (05). In excess of 150 compounds were identified from the breath samples provided.

**Table 3.** Volatile organic compounds identified in the breath of 8 diabetic children

| Compound name (CAS number) | Occurrence (max=40) | Occurrence (% of total samples) |
| --- | --- | --- |
| Acetaldehyde (75-07-0) | 40 | 100 |
| Acetic acid (64-19-7) | 18 | 45 |
| Acetone (67-64-1) | 40 | 100 |
| Allyl methyl sulfide (10152-76-8) | 40 | 100 |
| Benzene (71-43-2) | 40 | 100 |
| Butanal (123-72-8) | 40 | 100 |
| 1-butanol (71-36-3) | 40 | 100 |
| 2-butanol (78-92-2) | 40 | 100 |
| 2-butanone (78-93-3) | 40 | 100 |
| Cis-1-ethyl-2methylcyclohexane (4923-77-7) | 8 | 20 |
| Cyclohexane (110-82-7) | 40 | 100 |
| Decane (124-18-5) | 10 | 25 |
| 5,6 dimethyldecane (1636-43-7) | 14 | 35 |
| 2,3-dimethylheptane (3074-71-3) | 5 | 12.5 |
| 2,4-dimethylheptane ( 2213-23-2) | 40 | 100 |
| 2,4-dimethyl-1-heptene (19549-87-2) | 40 | 100 |
| (1S)-6,6-dimethyl-2-methylene-bicyclo[3.1.1]heptane (18172-67-3) | 6 | 15 |
| Dimethyl sulfone (67-71-0) | 40 | 100 |
| E,E-2,6-dimethyl-1,3,5,7-octatetraene (460-01-5) | 5 | 12.5 |
| Ethylbenzene (100-41-4) | 40 | 100 |
| 3-ethylheptane(2216-32-2) | 5 | 12.5 |
| 3-ethylhexane (619-99-8) | 8 | 20 |
| 1-ethyl-3-methylbenzene (620-14-4) | 36 | 90 |
| Heptanal (111-71-7) | 40 | 100 |
| Heptane (142-82-5) | 40 | 100 |
| 2-heptanone (110-43-0) | 40 | 100 |
| 1-heptene (592-76-7) | 23 | 57.5 |
| 1-hexadecyne (629-74-3) | 4 | 10 |
| 1,1,1,3,3,3-hexafluoro-2-propanol ( 920-66-1) | 4 | 10 |
| Hexanal (66-25-1) | 40 | 100 |
| Hexane (110-54-3) | 40 | 100 |
| 1-hexanol (111-27-3) | 4 | 10 |
| Isoprene (78-79-5) | 40 | 100 |
| Isopropyl alcohol (67-63-0) | 40 | 100 |
| Methacrolein (78-85-3) | 40 | 100 |
| Methanol (67-56-1) | 40 | 100 |
| 2-methylbutane (78-78-4) | 35 | 87.5 |
| 2-methyl-2-butanol (75-85-4) | 5 | 12.5 |
| 2-methyl-3-buten-2-ol (115-18-4) | 40 | 100 |
| 3-methyl-3-buten-1-ol (763-32-6) | 22 | 55 |
| 3-methylheptane ( 589-81-1) | 11 | 27.5 |
| 4-methylheptane ( 589-53-7) | 40 | 100 |
| 3-methylhexane (589-34-4) | 21 | 52.5 |
| 3-methyloctane (2216-33-3) | 10 | 25 |

| Compound | Count | Percent |
|---|---|---|
| 4-methyloctane (2216-34-4) | 40 | 100 |
| 3-methylpentane 996-14-0) | 4 | 10 |
| 2-methyl-1-pentene (763-29-1) | 40 | 100 |
| 2-methyl-1-propanol (78-83-1) | 38 | 95 |
| 1(methylthio) propane (3877-15-4) | 18 | 45 |
| 1-(methylthio)- (E)-1-propene (24848-60-4) | 32 | 80 |
| Methyl vinyl ketone (78-94-4) | 40 | 100 |
| m-xylene (108-38-3) | 40 | 100.0 |
| n-hexyl acrylate (2499-95-8) | 5 | 12.5 |
| Nonane (111-84-2) | 40 | 100 |
| 2,2,4,6,6-pentamethylheptane (13475-82-6) | 15 | 37.5 |
| Pentanal (110-62-3) | 40 | 100 |
| Pentane (109-66-0) | 35 | 87.5 |
| 2-pentanone (107-87-9) | 40 | 100 |
| Pentylcyclopropane (2511-91-3) | 12 | 30 |
| beta-pinene (127-91-3) | 40 | 100 |
| Propanoic acid (79-09-4) | 7 | 17.5 |
| 1-propanol (71-23-8) | 40 | 100 |
| 1-propoxy-2-propanol (1569-01-3) | 9 | 22.5 |
| Propylbenzene (103-65-1) | 16 | 40 |
| p-xylene (106-42-3) | 40 | 100 |
| Styrene (100-42-5) | 35 | 87.5 |
| Tetrachloroethylene (127-18-4) | 5 | 12.5 |
| Toluene(108-88-3) | 40 | 100 |
| 2,6,7-trimethyldecane (62108-25-2) | 7 | 17.5 |
| 2,3,5-trimethylhexane (1069-53-0) | 4 | 10 |
| Unknown (10.93 mins) | 11 | 27.5 |
| Unknown (RT 11.42) | 40 | 100 |
| Unknown RT 13.88 min | 11 | 27.5 |
| Unknown RT 14.71 minutes | 4 | 10 |

74 compounds were present in at least 10% of the patient samples (Table 3), 36 of these were found to be present in 100% of the patient samples. Phenol and N,N-dimethylacetamide were also found in all patient samples but are known contaminants of Tedlar® bags and are not presented in Table 3 [Steeghs *et al.* 2007]. The ubiquitous compounds contain a series of 2-substituted ketones including acetone, 2-butanone, 2-pentanone and 2-heptanone. Although 2-hexanone was detected it was not present in 10% of the patient samples. The statistical analysis focused on the 36 ubiquitous compounds. Table 4 shows the correlation coefficients when the peak area values of the 36 ubiquitous compounds were compared to the blood glucose levels for each patient at each time point. The overall correlation coefficient which compares all patient blood glucose levels with the 40 measured peak area values for that compound is also included. The correlation coefficients were calculated using the CORREL function in Microsoft Excel.

**Table 4.** The Correlation of the peak area of 36 VOCs with the individual patients blood glucose levels. The overall correlation between all measured peak area values for each VOC and the matched blood glucose measurements is also included.

| Compound | Correlation coefficient | | | | | | | | |
|---|---|---|---|---|---|---|---|---|---|
| | F1 | F2 | F3 | M1 | M2 | M3 | M4 | M5 | Overall |
| Acetaldehyde | -0.51 | -0.02 | -0.61 | 0.09 | 0.03 | 0.46 | -0.71 | 0.67 | 0.3 |
| Acetone | 0.05 | 0.62 | 0.31 | 0.92 | -0.38 | -0.62 | 0.43 | -0.39 | -0.11 |
| Allyl methyl sulphide | 0.31 | 0.33 | 0.20 | -0.40 | 0.30 | 0.71 | 0.55 | -0.69 | 0.29 |

| Compound | | | | | | | | | |
|---|---|---|---|---|---|---|---|---|---|
| Benzene | 0.42 | -0.50 | -0.05 | -0.35 | 0.21 | -0.03 | -0.58 | -0.05 | 0.28 |
| Butanal | -0.42 | -0.16 | -0.44 | -0.29 | 0.71 | 0.88 | -0.69 | 0.43 | 0.19 |
| 2-butanol | -0.29 | -0.61 | 0.16 | -0.04 | 0.77 | -0.38 | -0.74 | 0.07 | 0.01 |
| 2-butanone | -0.62 | -0.17 | 0.74 | -0.22 | 0.92 | -0.40 | -0.45 | -0.28 | -0.01 |
| Cyclohexane | 0.63 | 0.83 | -0.39 | -0.24 | 0.46 | -0.54 | -0.61 | -0.84 | -0.30 |
| 2,4-dimethyl-1-heptene | 0.39 | 0.05 | -0.67 | -0.40 | 0.35 | -0.23 | -0.45 | 0.18 | -0.24 |
| 2,4-dimethylheptane | 0.47 | -0.06 | -0.62 | -0.36 | 0.42 | -0.39 | -0.39 | 0.23 | -0.25 |
| Dimethyl sulfone | 0.19 | -0.50 | -0.26 | -0.34 | 0.03 | -0.11 | -0.88 | 0.37 | -0.08 |
| Ethylbenzene | 0.10 | 0.78 | -0.85 | -0.67 | 0.08 | 0.23 | -0.57 | -0.24 | -0.27 |
| 1-ethyl-3-methylbenzene | 0.80 | 0.88 | 0.73 | -0.47 | 0.92 | -0.38 | -0.24 | 0.07 | -0.29 |
| Heptanal | -0.54 | -0.18 | -0.33 | -0.28 | 0.72 | 0.19 | -0.55 | 0.54 | 0.19 |
| Heptane | 0.05 | 0.22 | -0.70 | -0.13 | 0.35 | -0.59 | -0.52 | 0.19 | -0.11 |
| 2-heptanone | -0.38 | -0.05 | -0.70 | -0.14 | 0.43 | 0.16 | -0.97 | 0.45 | 0.26 |
| Hexanal | -0.58 | -0.27 | -0.50 | -0.30 | 0.50 | 0.64 | -0.82 | 0.55 | 0.23 |
| Hexane | -0.02 | -0.03 | -0.77 | -0.12 | 0.42 | -0.58 | -0.30 | 0.28 | -0.19 |
| Isoprene | 0.04 | -0.82 | -0.34 | 0.12 | -0.08 | -0.86 | -0.76 | -0.34 | -0.49 |
| Isopropyl alcohol | -0.26 | -0.77 | -0.09 | 0.35 | -0.53 | -0.25 | 0.09 | -0.09 | -0.45 |
| Methacrolein | 0.43 | -0.71 | 0.17 | 0.95 | 0.17 | 0.16 | -0.50 | -0.27 | 0.37 |
| Methanol | -0.44 | -0.11 | 0.15 | -0.10 | 0.75 | 0.85 | -0.53 | 0.64 | -0.13 |
| 2-methyl-3-buten-2-ol | 0.11 | -0.35 | -0.32 | -0.12 | 0.25 | -0.70 | 0.52 | 0.11 | 0.22 |
| 4-methylheptane | 0.56 | 0.13 | -0.63 | -0.34 | 0.32 | -0.53 | -0.41 | 0.24 | -0.23 |
| 4-methyloctane | 0.55 | -0.05 | -0.61 | -0.43 | 0.36 | -0.25 | -0.39 | 0.00 | -0.27 |
| 2-methyl-1-pentene | 0.29 | 0.08 | -0.52 | -0.14 | 0.24 | -0.53 | -0.40 | 0.50 | -0.14 |
| Methyl vinyl ketone | -0.33 | 0.29 | -0.61 | -0.48 | 0.82 | -0.95 | -0.90 | -0.07 | -0.24 |
| m-xylene | -0.11 | 0.55 | -0.80 | -0.85 | -0.06 | 0.97 | -0.58 | -0.36 | -0.29 |
| Nonane | 0.57 | 0.29 | -0.79 | -0.43 | 0.45 | -0.50 | -0.85 | -0.61 | -.35 |
| Pentanal | -0.39 | 0.07 | -0.63 | -0.12 | 0.52 | -0.11 | -0.70 | 0.38 | 0.26 |
| 2-pentanone | -0.70 | 0.04 | 0.09 | -0.41 | 0.80 | -0.37 | -0.47 | 0.03 | -0.25 |
| beta-pinene | 0.38 | -0.75 | 0.39 | -0.56 | 0.73 | -0.08 | 0.36 | -0.60 | -0.17 |
| 1-propanol | -0.26 | -0.61 | 0.49 | -0.09 | 0.14 | -0.75 | -0.27 | 0.54 | 0.02 |
| p-xylene | -0.01 | 0.79 | -0.31 | -0.94 | 0.13 | 0.72 | -0.44 | -0.41 | -0.27 |
| Toluene | 0.79 | -0.03 | -0.81 | 0.27 | -0.16 | -0.45 | -0.89 | -0.43 | -0.41 |
| Unknown (RT 11.42) | -0.08 | 0.46 | -0.81 | -0.29 | 0.56 | -0.15 | -0.89 | -0.48 | -0.06 |

Table 4 shows that none of the ubiquitous compounds shows a significant overall correlation with blood glucose levels. Isoprene has the highest overall correlation with blood glucose, exhibiting a slight negative correlation of -0.49. Acetone and other ketones showed no overall correlation with blood glucose levels. Certain patients showed high individual correlations for specific compounds. However, there was not a consistent pattern of positive or negative individual correlations with respect to blood glucose for any of the compounds identified. If an individual correlation value of ≥0.75 (or -0.75) is taken as being indicative of a significant relationship between blood glucose levels and the peak area of the compound, then the following compounds exhibited no significant correlation with the individual patients blood glucose levels: acetaldehyde, 2-methyl-1pentene, 2-methyl-3-buten-2-ol, benzene, heptane, allyl methyl sulphide, pentanal, 4-methylheptane, 2,4-dimethylheptane, 2,4-dimethyl-1-heptene, 4-methyloctane and heptanal. This list of compounds contains none of the ketones identified but 3/5 of the aldehydes and 5/6 of the straight chain or branched alkanes. In addition it also contains allyl methyl sulphide, a volatile usually associated with the ingestion of garlic. Isoprene has a negative correlation with blood glucose levels in 3 of the 8 patients (M4, M3 and F2). 1-ethyl-3-methylbenzene has a positive correlation with blood glucose levels for 3 of the 8 patients (M2, F2 and F1) and patient F3 has a positive correlation just below the threshold. Methanol has a positive correlation with blood glucose for 2 of the patients (M2 and M3). Nonane shows a negative correlation with blood glucose for 2 of the patients (F3 and M4). Of the ketones acetone (M1), 2-butanone (M2) and 2-pentanone (M2) show a positive correlation in only 1 patient. In contrast 2-heptanone exhibits a negative correlation with blood glucose in patient M4. There were some compounds that showed both negative and positive correlations with blood glucose levels for different patients. These compounds include methyl vinyl ketone, toluene, m-xylene, cyclohexane, ethylbenzene and p-xylene. These are mainly volatiles that might be associated with an exogenous source such as the environment.

3.3 *SIFT-MS*
The SIFT-MS results for 24 compounds are displayed in Table 5. A control is included who had well controlled blood glucose spanning a small range (mean 4.5, s.d. 0.6) when compared with the patient group (mean 12.2, s.d. 4.98). The compounds present in the largest abundance were acetone, ammonia, nitric oxide, β-pinene, isoprene and the alcohols methanol, ethanol and 1-propanol. The mean overall concentrations of these compounds including both control and patient data were, acetone (279ppb), ammonia (243ppb), nitric oxide (365ppb), β-pinene (39.76ppb), isoprene (28.50ppb), methanol (72ppb), ethanol (457ppb) and 1-propanol (96ppb). This compares well with literature values for acetone and the other compounds measured using SIFT-MS. A large study of 243 individuals using PTRMS [Schwarz *et al.* 2009] found that acetone concentrations ranged between 281-1246ppb with a geometric mean of 544ppb. The same study considered a group of 44 children (5-11years) whose median breath acetone level was 263ppb. Another study found that the mean breath acetone concentration for young adults and children was 263ppb whereas for adults aged 20-60 it was significantly higher at 477ppb [Spanel *et al.* 2007]. The same paper found that the average ammonia level in young adults was 317 ppb rising to 833ppb in adults. A study of breath isoprene in healthy volunteers [Turner et al. 2006] found the mean level was 118ppb with a range of (0-474ppb). However, a study by Smith *et al.* [Smith *et al.* 2009] focusing on breath isoprene levels in 200 healthy children found that the mean levels varied with age (7–10 years (28 ± 24 ppb), 10–13 years (40 ± 21 ppb), 13–16 years (60 ± 41 ppb)) and were generally lower than the adult population. The distribution of the children's data were compared with data from an adult cohort also measured by SIFT-MS and the median levels were 37ppb and 106ppb respectively. This data is in agreement with the data measured in this study involving diabetic children in the age range 7-16. The mean levels of ethanol (mean 196ppb, range 0-1663ppb) and acetaldehyde (mean 24ppb, range 0-104ppb) were measured in 30 volunteers using SIFT MS [Turner et al. 2006b].

A study by Turner *et al.* [2009] involving 8 adults with type 1 diabetes found much higher average levels of acetone than identified in this study. However, this fits with the age-related results discussed above which highlight an increase in the mean breath acetone concentration with age. It should also be noted that the diabetic patients in the study of Turner *et al.* had a relatively long duration of disease (mean 28± 3 years) compared to this study involving children (mean 5 years). It is interesting that the patients in this study show consistently low breath acetone levels despite their blood glucose levels being over a large range (122.4-435.6mg/dL). This is compared to the study of Turner *et al.* where the average breath acetone concentration prior to insulin clamp was 5.5ppm and the range of baseline blood glucose values was 86.4-250.2md/dL. It is possible that the observed differences are partly age-related but could also be associated with the longevity of disease. It is likely that disease duration and history of glycaemic control play a major role in altering the "normal" metabolism. This highlights a problem with comparing the outcomes of breath volatile studies where cohorts are made up of different age ranges. It also shows that age, gender, disease history etc. would have to be factored into any developed test. However it should be noted that even though the mean/median values obtained from these cohort studies often show differentiation the ranges often overlap considerably. This type of compensation could potentially prove problematic where such wide individual variation exists within groups.

The other ketones 2-butanone, 2-pentanone and 2-heptanone also observed in the ATD study are present at very low mean concentrations, at 0.8ppb, 0.93ppb and 1.34ppb respectively. This is in agreement with the ATD results where the peak area of acetone was 2-3 orders of magnitude greater than the peak area of the other ketones.

There was no significant overall correlation between the concentration of any of the compounds and the blood glucose levels. The highest overall correlation was observed for acetone (-0.64). However, this negative correlation predominantly arises because the acetone concentrations are significantly higher for the control than the patients. If the patients are treated as an individual group then no overall correlation exists, in agreement with the GCMS study. The mean value of acetone is 523 ppb for the control but only 198ppb for the patient group. This can be compared with isoprene where the mean levels are lower in the control (21.78ppb) than the patient group (30.74ppb), and also acetaldehyde where the mean values are very well matched (control 27.28ppb, patient group 26.58ppb). The mean blood glucose levels for the patients (219.6mg/dL) are higher than the control (81mg/dL) and the ranges do not overlap. Therefore, the results may simply reflect "normal metabolism" where ketone levels would be expected to be higher when blood glucose levels were lower. It is not possible to be certain of the effect of low blood glucose without testing the patients over a lower range of values using for example an insulin clamp regime to identify a definite link to metabolism or indeed to identify a difference between adults with type 1 diabetes.

The mean acetone levels were found to be higher at the beginning of the day (340ppb) versus the end of the session (168ppb). This is a consistent observation for the control and patient groups and appears to be generally independent of blood glucose level (mean at beginning of day 221.4mg/dL versus the mean at the end of the session 244.8mg/dL). A recent study by Wang *et al.* [Wang CJ *et al.* 2010] of adults with type 1 diabetes and an earlier study by Tassopoulos *et al.* [1969] found that acetone levels and blood glucose followed a cyclical pattern during the day. Tassopoulos *et al.* found that breath acetone and blood glucose were high in the morning and reached their lowest levels in the late afternoon. This agrees with our observations for breath acetone concentrations. However, we did not identify a trend with blood glucose. However, Wang *et al.* did identify that breath acetone peaks were often delayed with respect to peaks in blood glucose levels with differences of up to 4 hours observed for some patients. Therefore, a longer study in terms of time measurements and including more patients may help establish a reliable trend. However, this complexity just highlights further the difficulty in linking any compounds with blood glucose levels. Presumably this differential time aspect exists for many compounds and varies depending on the metabolic pathway involved.

Some of the subjects showed individual correlation with one or more of the VOCs and blood glucose (see Table 6). Patient F1 again showed a positive correlation between toluene concentration and blood glucose levels in agreement with the ATD study. However, positive correlations were also observed for acetic acid, 2-butanone, 1-butanol, heptanal, nonane, 2-pentanone, and p-xylene. Patient M5 again exhibited limited correlation with volatiles and blood glucose levels in agreement with the ATD study.

|  | Control A | | | | | F1 | | | | | M4 | | | | | M5 | | | | |
| --- | --- | --- | --- | --- | --- | --- | --- | --- | --- | --- | --- | --- | --- | --- | --- | --- | --- | --- | --- | --- |
|  | Sample 01 | Sample 02 | Sample 03 | Sample 04 | Sample 05 | Sample 01 | Sample 02 | Sample 03 | Sample 04 | Sample 05 | Sample 01 | Sample 02 | Sample 03 | Sample 04 | Sample 05 | Sample 01 | Sample 02 | Sample 03 | Sample 04 | Sample 05 |
| Blood glucose (mg/dL) | 70.2 | 68.4 | 91.8 | 86.4 | 88.2 | 180 | 288 | 309.6 | 189 | 154.8 | 340.2 | 226.8 | 270.4 | 361.8 | 300.6 | 297 | 226.8 | 122.4 | 360 | 435.6 |
| Compound (CAS number) | | | | | | | | | | Concentration ppb | | | | | | | | | | |
| acetaldehyde (75-07-0) | 27.18 | 25.53 | 27.62 | 19.71 | 36.34 | 36.17 | 31.18 | 33.43 | 21.82 | 17.21 | 28.23 | 24.93 | 27.06 | 24.36 | 13.57 | 31.57 | 30.32 | 29.98 | 20.65 | 28.16 |
| acetic acid (64-19-7) | 13.40 | 14.21 | 13.24 | 9.35 | 9.28 | 8.47 | 8.38 | 9.44 | 6.76 | 7.20 | 9.67 | 7.93 | 8.64 | 5.49 | 5.38 | 10.41 | 10.28 | 9.04 | 7.64 | 8.75 |
| acetone (67-64-1) | 577.26 | 706.59 | 617.91 | 375.43 | 338.00 | 194.72 | 162.36 | 149.16 | 164.73 | 85.31 | 368.26 | 276.24 | 294.16 | 165.39 | 108.54 | 280.37 | 231.17 | 219.94 | 131.35 | 139.89 |
| ammonia (7664-41-7) | 648.85 | 333.93 | 110.48 | 576.94 | 88.38 | 190.82 | 223.70 | 526.50 | 620.50 | 227.60 | 141.78 | 149.12 | 108.04 | 80.59 | 65.38 | 188.63 | 143.19 | 123.03 | 120.65 | 201.25 |
| benzene (71-43-2) | 3.86 | 2.32 | 2.91 | 3.21 | 2.77 | 2.86 | 2.59 | 3.16 | 2.57 | 2.02 | 3.42 | 2.33 | 2.08 | 1.42 | 1.54 | 3.22 | 2.51 | 2.61 | 2.09 | 2.67 |
| beta-pinene (127-91-3) | 24.14 | 28.81 | 40.23 | 40.71 | 99.81 | 34.94 | 95.85 | 47.35 | 21.05 | 22.41 | 46.28 | 30.17 | 33.43 | 22.34 | 19.25 | 38.99 | 61.79 | 39.59 | 21.04 | 26.97 |
| butanoic acid (107-92-6) | 8.93 | 8.79 | 8.65 | 6.61 | 8.52 | 4.97 | 5.22 | 6.84 | 4.77 | 5.37 | 6.74 | 4.65 | 5.29 | 3.69 | 3.00 | 6.18 | 8.82 | 5.41 | 4.87 | 7.06 |
| 1-butanol (71-36-3) | 1.37 | 0.96 | 1.20 | 0.87 | 0.81 | 0.95 | 0.97 | 1.07 | 0.72 | 0.74 | 0.87 | 0.95 | 0.80 | 0.54 | 0.56 | 1.06 | 1.13 | 0.84 | 0.78 | 0.55 |
| butanone (78-93-3) | 0.96 | 1.46 | 1.01 | 0.58 | 1.24 | 0.65 | 0.76 | 0.85 | 0.55 | 0.56 | 0.78 | 0.96 | 0.71 | 0.50 | 0.35 | 0.95 | 1.31 | 0.73 | 0.60 | 0.54 |
| ethanol (64-17-5) | 250.72 | 347.40 | 485.29 | 205.68 | 1214.78 | 429.55 | 464.19 | 1118.53 | 150.24 | 497.86 | 348.26 | 172.24 | 266.23 | 179.16 | 714.24 | 491.30 | 406.07 | 422.26 | 192.38 | 779.03 |
| heptanal (111-71-7) | 2.39 | 2.44 | 2.05 | 2.08 | 1.43 | 2.07 | 2.06 | 2.37 | 1.44 | 1.14 | 2.11 | 2.35 | 1.99 | 1.46 | 1.04 | 2.61 | 2.70 | 2.00 | 1.90 | 1.03 |
| heptane (142-82-5) | 18.30 | 18.06 | 22.06 | 11.55 | 59.11 | 13.04 | 13.57 | 10.89 | 11.63 | 11.23 | 20.93 | 10.85 | 14.15 | 11.59 | 9.40 | 17.71 | 14.00 | 12.58 | 8.15 | 35.69 |
| 2-heptanone (110-43-0) | 1.75 | 1.46 | 1.63 | 1.41 | 1.47 | 1.43 | 1.47 | 1.27 | 1.73 | 1.18 | 1.36 | 1.15 | 1.16 | 1.05 | 1.18 | 1.31 | 1.47 | 1.17 | 1.26 | 0.97 |
| hexanal (66-25-1) | 2.76 | 2.19 | 1.71 | 1.58 | 1.32 | 2.41 | 2.31 | 2.05 | 1.47 | 0.93 | 2.05 | 1.89 | 1.60 | 1.28 | 0.99 | 2.07 | 2.00 | 2.07 | 1.32 | 0.99 |
| hexane (110-54-3) | 8.48 | 15.49 | 17.06 | 6.00 | 15.79 | 7.68 | 7.88 | 14.57 | 7.64 | 8.10 | 7.28 | 7.16 | 6.57 | 6.49 | 11.04 | 9.93 | 9.03 | 8.82 | 4.86 | 11.66 |
| isoprene (78-79-5) | 30.29 | 19.70 | 22.73 | 24.06 | 12.12 | 45.12 | 57.95 | 56.78 | 27.10 | 33.01 | 37.88 | 21.66 | 45.22 | 12.55 | 6.96 | 23.72 | 31.75 | 38.59 | 14.85 | 8.05 |
| methanol (67-56-1) | 149.31 | 136.51 | 105.27 | 88.12 | 47.35 | 75.83 | 64.18 | 67.99 | 54.68 | 29.39 | 40.29 | 41.74 | 49.78 | 43.21 | 27.85 | 130.26 | 101.38 | 82.20 | 62.56 | 32.88 |
| nitric oxide (10102-43-9) | 369.32 | 447.10 | 431.02 | 408.94 | 259.06 | 346.77 | 451.21 | 426.60 | 330.48 | 220.48 | 442.63 | 431.58 | 430.53 | 284.90 | 210.60 | 448.02 | 444.27 | 414.32 | 306.62 | 206.70 |
| nonane (111-84-2) | 5.98 | 5.30 | 5.37 | 4.02 | 4.62 | 4.04 | 5.01 | 5.33 | 4.66 | 3.86 | 4.81 | 4.73 | 4.61 | 3.60 | 3.76 | 5.27 | 4.46 | 5.26 | 4.03 | 4.94 |
| 2-pentanone (107-87-9) | 1.38 | 1.81 | 1.20 | 0.81 | 1.01 | 0.82 | 0.86 | 0.96 | 0.67 | 0.48 | 0.97 | 1.28 | 0.97 | 0.71 | 0.58 | 0.97 | 1.20 | 0.68 | 0.72 | 0.50 |
| propane (74-98-6) | 14.71 | 10.67 | 8.30 | 5.76 | 5.84 | 60.21 | 10.73 | 9.55 | 6.26 | 5.79 | 17.73 | 7.59 | 7.78 | 5.97 | 6.26 | 30.03 | 8.54 | 7.12 | 2.94 | 6.05 |
| 1-propanol (71-23-8) | 75.11 | 94.50 | 163.45 | 69.21 | 126.33 | 101.32 | 131.09 | 104.52 | 47.88 | 103.38 | 71.80 | 48.14 | 63.81 | 65.93 | 70.24 | 138.40 | 126.53 | 128.11 | 68.86 | 128.51 |
| p-xylene (106-42-3) | 0.51 | 0.49 | 0.52 | 0.47 | 0.90 | 0.38 | 0.62 | 0.66 | 0.41 | 0.36 | 0.35 | 0.43 | 0.49 | 0.29 | 0.33 | 0.58 | 0.66 | 0.48 | 0.54 | 0.49 |
| toluene (108-88-3) | 1.66 | 1.44 | 1.08 | 1.43 | 0.62 | 1.11 | 1.79 | 1.83 | 0.78 | 0.75 | 1.17 | 1.33 | 1.32 | 0.86 | 0.68 | 1.30 | 1.43 | 1.33 | 0.94 | 0.75 |

**Table 5** SIFT-MS data for 3 patients with type 1 diabetes and a healthy control. The concentration (ppb) of 24 breath volatiles versus the corresponding blood glucose concentration (mg/dL).

**Table 6** showing the correlation of compounds detected by SIFT-MS and their relationship to blood glucose

| Compound | Correlation coefficient | | | | |
|---|---|---|---|---|---|
| | ConA | F1 | M5 | M4 | Overall |
| acetaldehyde | 0.20 | 0.57 | -0.45 | -0.15 | -0.06 |
| acetic acid | -0.61 | 0.76 | -0.38 | -0.36 | -0.54 |
| acetone | -0.58 | 0.24 | -0.60 | -0.26 | -0.64 |
| ammonia | -0.56 | 0.23 | 0.59 | -0.30 | -0.32 |
| benzene | -0.09 | 0.65 | -0.09 | 0.00 | -0.28 |
| beta-pinene | 0.56 | 0.74 | -0.60 | -0.06 | -0.13 |
| butanoic acid | -0.41 | 0.68 | 0.00 | -0.11 | -0.46 |
| 1-butanol | -0.33 | 0.81 | -0.56 | -0.52 | -0.48 |
| butanone | -0.40 | 0.93 | -0.47 | -0.46 | -0.46 |
| ethanol | 0.45 | 0.67 | 0.37 | 0.16 | 0.09 |
| heptanal | -0.74 | 0.79 | -0.58 | -0.46 | -0.25 |
| heptane | 0.36 | 0.08 | 0.59 | 0.21 | -0.14 |
| 2-heptanone | -0.25 | -0.09 | -0.45 | 0.11 | -0.59 |
| hexanal | -0.84 | 0.57 | -0.86 | -0.21 | -0.32 |
| hexane | 0.20 | 0.68 | 0.11 | 0.14 | -0.28 |
| isoprene | -0.39 | 0.85 | -0.99 | -0.48 | -0.09 |
| methanol | -0.78 | 0.51 | -0.53 | -0.41 | -0.50 |
| nitric oxide | -0.27 | 0.90 | -0.79 | -0.46 | -0.25 |
| nonane | -0.56 | 0.93 | -0.33 | -0.52 | 0.53 |
| 2-pentanone | -0.79 | 0.83 | -0.40 | -0.55 | -0.48 |
| propane | -0.80 | -0.27 | -0.07 | 0.28 | -0.04 |
| 1-propanol | 0.61 | 0.46 | -0.30 | 0.58 | -0.09 |
| p-xylene | 0.37 | 1.00 | -0.12 | -0.97 | -0.21 |
| toluene | -0.69 | 0.96 | -0.85 | -0.65 | -0.17 |

*3.4 General discussion*

A recent perspective article gives a very good summary of the state of the art with respect to breath analysis and monitoring of diabetes [Smith *et al.* 2011].

Acetone is formed via the decarboxylation of acetoacetate and the dehydrogenation of isopropylalcohol [Kalapos 2003]. Acetone and other ketone bodies are formed by the liver during the metabolism of fatty acids. In the absence of readily available glucose this pathway may predominate. Therefore, insulin dependent diabetics with poor glycaemic control are susceptible to ketoacidosis a condition where high levels of ketone bodies (acetone, acetoacetate and β-hydroxybutyrate) are present in the blood and excreted in the urine. Breath acetone levels have been reported to be higher in diabetes [Deng *et al.* 2004 and Nelson *et al.* 1998]. For example Deng *et al.* found that the median acetone concentration for healthy controls was 520ppb and for type 2 diabetics it was 2260ppb whereas Nelson *et al.* found that diabetic children had significantly higher levels of acetone than healthy controls. However, in healthy individuals and controlled diabetics the generation of ketones represents the normal physiological response to lower concentrations of blood glucose – "starvation ketones". Thus the concentration of breath acetone has been previously reported to be inversely related to the concentration of blood glucose in healthy volunteers [Turner *et al.* 2008]. There is little information available on breath acetone concentrations for diabetic patients in a non-ketotic state. This study tested type I diabetics in the non-fasting state and found no overall relationship between blood glucose and acetone concentrations and is consistent with other studies where the baseline levels of acetone for each patient varied widely.

A number of studies have been undertaken to establish links between breath volatiles and diabetes. Turner *et al.* [Turner *et al.* 2009] used SIFT-MS to study VOCs in 8 adults with type 1 diabetes patients tested under strict hypoglycaemic control. Each patient demonstrated a positive relationship between acetone and blood glucose concentrations, although the range of acetone levels varied widely between patients. In contrast this study involved children in a non-fasting state who had a wide range of blood glucose readings. Wang [Wang C J *et al.* 2010] studied 34 patients with type 1 diabetes and found no overall correlation with breath acetone. However, if the patients were grouped into 4 distinct classes according to their blood glucose levels and the mean plotted then a positive relationship with breath acetone was found. This study is in partial agreement with these results as it was observed that there was no overall correlation when individual breath acetone results were compared with blood glucose measurements.

Isoprene is commonly found in breath. There are several routes by which it is generated including the mevalonic acid pathway of cholesterol [Buszewski *et al.* 2007]. It is known that poor diabetic control is associated with dyslipidaemia but this is slow to emerge [Hallikainen *et al.* 2006]. A study found there was no relationship between the concentration of exhaled breath isoprene and serum cholesterol concentration [Buszewski *et al.* 2007]. Breath isoprene is influenced by sex (it is higher in males) [Lechner *et al.* 2006] and was found to be positively associated with age [Nelson *et al.* 1998]. Two studies have addressed the relationship between breath isoprene and blood glucose [Turner *et al.* 2008 and Nelson *et al.* 1998]. Nelson *et al.* reported no difference between the amount of isoprene in exhaled breath in healthy controls compared with those with diabetes, or between the level in diabetics in a fasting or non-fasting/post insulin state. However, the 8 diabetic children studied were only sampled once before and after food and no attempt was made to quantify any change in blood glucose level, or to correlate this with changes in isoprene levels. Turner *et al.* found no relationship between serum glucose and breath isoprene concentrations in healthy subjects. The range of blood glucose levels prior to ingestion of glucose, was within the range 72–108 mg/dL, rising 2-fold after ingestion. This study identified a slight overall negative correlation between breath isoprene and blood glucose levels in 8 diabetic children using ATD coupled with GC-MS. The SIFT-MS data did confirm a slight negative correlation (for the patient group) with isoprene concentration and blood glucose level. However, if the patient's individual correlations are considered then the picture is less clear with strong negative and positive correlations observed between isoprene concentration and blood glucose. It is evident that a larger study is required to clarify these findings with respect to isoprene.

Oxidative stress plays a central role in the onset of diabetes mellitus and the subsequent development of vascular and neurological complications. It also likely to be a mechanism causing changes in the exhaled VOCs measured in diabetic patients. Exhaled VOCs (C4–C20 n-alkanes and their monomethylated derivatives) have been studied in patients with type 1 and 2 diabetes mellitus [Phillips *et al.* 2004], showing that oxidative stress was significantly and similarly increased in type 1 and type 2 diabetes mellitus but this increase appeared to be independent of glycaemic control.

Methyl nitrate production is related to oxidative stress, which is regulated by insulin through its modulation of lipolysis. Methyl nitrate was reported to be correlated with blood glucose in children with type 1 diabetes [Novak et al. 2007]. Approximately 100 exhaled gas species were defined and compared with plasma glucose concentrations and the VOC displaying the greatest correlation with blood glucose levels was methyl nitrate. Exhaled methyl nitrate concentrations closely matched plasma glucose levels in 16 out of the 18 subjects studied. The characteristics of methyl nitrate formation suggest that rather than reflecting hyperglycaemia per se, it may actually reflect the specific and complex pattern of metabolic alteration that accompanies hyperglycaemia in type 1 diabetes mellitus. A further study by Lee *et al.* [Lee *et al.* 2009] also showed a relationship between methyl nitrate concentration and hyperglycaemia in young adults with type 1 diabetes. Our study failed to identify methyl nitrate in the breath of diabetic children. Despite using pre-concentration the levels of methyl nitrate (low ppt) are probably too low and require specialist methodology and detectors. We were able to detect a standard of methyl nitrate at a higher concentration using GCMS so it is just a problem with the limit of detection. Lee *et al.* [Lee *et al.* 2009] also found that acetone, ethanol, xylene and ethylbenzene are correlated with hyperglycaemia. Exhaled ethanol is associated with fermentation by gut flora and would be expected to be affected by an increased hyperglycaemic state. It is assumed that aromatic hydrocarbons have an environmental source but it is possible that levels are increased in the breath due to changes in liver metabolism caused by hyperglycaemia. This study did not identify any correlation with ethanol concentration on the breath of diabetic children. Levels were very variable for all the subjects tested using SIFT-MS which could reflect changes due to consumption of food and beverages. Individual correlations for a number of aromatic hydrocarbons were identified, 1-ethyl-3-methylbenzene in particular showed a positive correlation with blood glucose levels in 3 patients. In addition other patients were identified with individual correlations to xylene, ethyl benzene and toluene although there were both negative and positive correlations within this patient group.

This study was designed to identify markers or groups of markers that could be used to provide a non-invasive basis for monitoring blood glucose levels. Despite many candidate markers or groups of markers being identified previously, not least acetone, this study failed to identify any definite correlation with these VOCs. The premise of this approach may be accurate: volatile compounds may correlate with blood glucose. However, the target compounds may be ill-advised. Ketones clearly rise in poorly controlled diabetes, but also occur in starvation states, so a bell-shaped relationship may exist between ketones and glucose; if this is true, then it is not surprising that linear modelling has failed to find an association. The second flaw may be timing: after meals, diabetics may have a 'glucose excursion' a rapid rise and, hopefully a fall, in glucose that is too short-lived for ketones to be produced in response to the apparent insulin insufficiency. In contrast, sick diabetic patients may be profoundly ketotic, without their blood sugar being especially high; ketosis is not proportional to blood glucose.

The results do provide detailed information on a number of compounds and their correlation (or lack thereof) with blood glucose levels over time. The results also give a comparison between two methods, GCMS and SIFT-MS which are in general agreement with regard to the main findings. Future work will aim to recruit larger numbers of patients and group them according to age and glycaemic control (current and historical).

**4. Conclusions**

The breath of 8 children with Type 1 diabetes was analysed using GCMS. A total of five breath samples and matched blood samples for glucose measurements were taken over a 6 hour period under hospital conditions. A total of 36 compounds was found to be present in all patient samples analysed. None of these compounds was found to exhibit a significant overall correlation with blood glucose. Isoprene showed a slight negative correlation overall with blood glucose levels. Some compounds exhibited significant individual correlations with the blood glucose levels of certain patients. However, there was no consistent pattern observed within these results which would enable a sound discriminatory model to be developed.

A SIFT-MS study was also undertaken on a subset of the patients and a healthy control. The levels of 24 compounds were monitored over time and compared with matched blood glucose levels. Again no compound showed a significant overall correlation with blood glucose levels. Acetone was found to be negatively correlated with blood glucose levels. However, this result was affected by the fact that the control had higher mean levels of acetone and consistently lower blood glucose. If the patient group were considered separately then no overall correlation was observed. The levels of acetone measured for diabetic patients (mean 198ppb, S.D. 78ppb) are lower than generally reported in previous studies, although previous work predominantly involved the study of adults with type 1 diabetes. However, this result is in agreement with previous SIFT-MS and PTR-MS studies that found that the mean breath acetone levels increase with age. Isoprene was again slightly negatively correlated with blood glucose levels if just the patient data was considered. Although this study failed to identify combinations of VOCs that may be useful for monitoring blood glucose levels, a more extensive study of diabetic patients from specific age ranges and graded according to their level of glycaemic control may facilitate firmer conclusions with respect to the viability of breath testing in this application.

**Acknowledgments**
The authors would like to acknowledge Diabetes UK for funding part of this work, the Bristol Royal Hospital for Children and the University of the West of England for the funding of a PhD studentship for S. Stevens.

**Disclaimer**
This project was funded by Diabetes UK and JPS is supported by the National Institute for Health Research Bristol Nutrition Biomedical Research Unit. The views expressed in this publication are those of the author(s) and not necessarily those of the NHS, the National Institute for Health Research or the Department of Health.